\documentclass[a4paper, 12pt,twoside, dvips] {article}

\usepackage[dvips]{graphicx}
\usepackage{epsfig}
\usepackage{amssymb}

\renewcommand{\baselinestretch}{1.2}

\begin{document}

{\normalsize\sf
\rightline {hep-ph/0203168}
\rightline{IFT-02/11}
\vskip 3mm
\rm\rightline{March 2002}
}
\vskip 5mm

\begin{center}
  
{\LARGE\bf Neutrino masses and unification of the gauge and Yukawa couplings}

\vskip 10mm

\setcounter{footnote}{-1}

{\large\bf Kamila Kowalska\footnote{email: Kamila.Kowalska@fuw.edu.pl }}\\[5mm]

Institute of Theoretical Physics, Warsaw University\\
Ho\.za 69, 00-681 Warsaw, Poland

\end{center}

\vskip 5mm

\renewcommand{\baselinestretch}{1.1} 
\begin{abstract}
\vskip 3mm 
There is a convincing experimental evidence that neutrinos are massive. Therefore we investigate the influence of the neutrino masses on the unification of gauge and Yukawa couplings in the framework of the Minimal Supersymmetric Standard Model. We estimate the contribution of the neutrino Yukawa coupling to the gauge and Yukawa coupling unification. We find that in the case of the gauge coupling unification the effect of massive neutrinos is small and can be neglected. It appears to be  much more significant, if we explore $Y_{b}$ and $Y_{\tau}$ equality at the GUT scale. The neutrino contribution can change that relation even by $\sim$12\%. 
\end{abstract}
\renewcommand{\baselinestretch}{1.2}

\section{Introduction}
The idea of Grand Unification~\cite{unif} has been known in theoretical physics for years. One of its most important predictions is the gauge couplings unification at the GUT scale. In such a framework the parameters of the Standard Model (or of its supersymmetric version) are related to the parameters of the unified theory by means of the renormalization-group evolution from the assumed GUT scale down to the electroweak scale. Conversely, the idea of unification of the strong and electroweak forces can be effectively tested by extrapolating the three known gauge couplings of the SM up to the energy scale $M_{GUT}$. Similarly, since some of the GUTs predict that the Yukawa couplings should also unify, the extrapolation of these couplings to the GUT scale provides further test of specific unification scenarios. Requiring unification may in this case put some limits on the low energy parameters. In the Minimal Supersymmetric Standard Model exact $b-\tau$ Yukawa coupling unification at the level of two-loop renormalization group equations is possible only for very small and very large values of $\tan\beta$, i.e. for $\tan\beta\lesssim 1.7$ and $\tan\beta\gtrsim 50$. Including supersymmetric one-loop correction one can enlarge the range of possible $\tan\beta$ values even to $\tan\beta\lesssim 2.1$ and $\tan\beta\gtrsim 10$~\cite{susycor}.

The existence of any additional field in a theory has important consequences in the context of the unification. The renormalization-group equations are sensitive to the particle contents of the theory under study. For example, the precise gauge coupling unification does not hold in the Standard Model~\cite{unsm}, while it becomes possible in the MSSM, where the superpartners of the ordinary particles are introduced. 

Recent data, coming from the experiments measuring the neutrino oscillations (\cite{sk}, \cite{chooz}, \cite{solar}, \cite{vacos},\cite{msw}) indicate that neutrinos have masses, yet still much smaller than the other fermions. In that connection, the Standard Model (or the MSSM) should be enlarged by the additional fields, describing the right-handed neutrinos. The most natural explanation of the smallness of the observed neutrino masses is to add $\nu_{R}$ with the mass $M_{R}\gg M_{Z}$ to the particle spectrum. This is always possible in the framework of the SM or MSSM, since right-handed neutrinos are singlets under gauge group $SU(3)\times SU(2)\times U(1)$. The effective neutrino mass is then generated through the see-saw mechanism~\cite{see-saw} 
\begin{equation}
m_{\nu}\sim\frac{M_{Z}^{2}}{M_{R}}.
\end{equation}
Obtaining light (i.e. $m_{\nu}\sim 1$ eV) neutrinos requires therefore the new scale $M_{R}$ to be very large, of the order of $10^{14}$ GeV. 

The influence of the neutrino masses on the unification of the gauge couplings in the MSSM has been studied in paper of J. A. Casas \textit{et al}~\cite{casas}. In the case of $b-\tau$ unification similar analysis for small values of $\tan\beta$ (without supersymmetric one-loop correction) has been performed by M. Carena \textit{et al}~\cite{Carena}. They found that the presence of massive neutrinos above the Majorana scale may enlarge the range of $\tan\beta$ consistent with the unification of the third generation Yukawa couplings, if there is large mixing in leptonic sector.

Encouraged by those results, in this paper we estimate the contribution of the neutrino Yukawa couplings to the running of gauge and Yukawa couplings for both small and large $\tan\beta$ regimes, however we do not consider mixing in leptonic sector. We consider different neutrino mass hierarchies as well as different values of the Majorana scale and of the superparticle decoupling scale $M_{SUSY}$. We also take into account the finite supersymmetric one-loop correction to the bottom quark mass. We conclude that the presence of the right-handed neutrinos above $M_{R}$ can change $Y_{\tau}/Y_{b}$ ratio even by $\sim$12\% and our results agree with those presented in~\cite{Carena} for small $\tan\beta$ and with no mixing. This effect can be used to obtain $b-\tau$ unification for other values of the MSSM parameters, like superparticle decoupling scale $M_{SUSY}$ or $\tan\beta$. However, the presence of massive neutrinos does not enlarge the range of $\tan\beta$ values, consistent with the $b-\tau$ unification. 

\section{Numerical results}
All calculations are performed by means of the two-loop renormalization-group equations for the gauge and Yukawa couplings. Below the scale $M_{SUSY}$ we use the Standard Model $\beta$-functions of the gauge and Yukawa couplings \cite{smrge}. Between $M_{SUSY}$ and $M_{R}$ the RGEs of MSSM \cite{rge} are used. Above the Majorana scale $M_{R}$, running of the neutrino Yukawa coupling must be taken into account. The appropriate two-loop RGEs have been derived using the general expression for the two-loop $\beta$ functions of the superpotential parameter $Y_{i}$ given in~\cite{rge}. We present them in the Appendix (see also \cite{ratz}).

We use the following input parameters :
\begin{eqnarray}
m_{t}(M_{t})=165\; GeV\qquad m_{b}(M_{b})=4.69\; GeV\qquad m_{\tau}=1.78\; GeV \nonumber\\
\alpha_{3}(M_{Z})=0.1185\qquad \alpha_{2}(M_{Z})=0.0338\qquad \alpha_{1}(M_{Z})=0.0169.\label{input} 
\end{eqnarray}
The initial values of the Yukawa couplings are obtained from the relations 
\begin{eqnarray}
Y_{t}(M_{t})=1.031\sqrt{1+\frac{1}{\tan^{2}\beta}}, \qquad Y_{b}(M_{b})=0.018\sqrt{1+\tan^{2}\beta}, \nonumber\\ 
Y_{\tau}=0.01\sqrt{1+\tan^{2}\beta}.\qquad\qquad\qquad\qquad\qquad\qquad\qquad\qquad\quad\;\label{yukin}
\end{eqnarray}

Form the oscillation experiments only neutrino mass-squared splittings can be obtained, not the masses themselves. Therefore, different assumptions about neutrino mass hierarchy can explain the experimental data: degenerate neutrino eigenstates $m_{1}\approx{m_{2}}\approx{m_{3}}$ with masses of order 1 eV or large hierarchy of neutrino masses $m_{1}\gg{m_{2}}\gg{m_{3}}$ or $m_{1}\gg m_{2}\approx{m_{3}}$. In this connection we consider such different sets of effective neutrino masses:

\begin{itemize}
\item $m_{\nu_{\tau}}=m_{\nu_{\mu}}=m_{\nu_{e}}=1$ eV
\item $m_{\nu_{\tau}}=4.5\cdot 10^{-2}$ eV, $m_{\nu_{\mu}}=10^{-3}$ eV, $m_{\nu_{e}}=10^{-4}$ eV
\item $m_{\nu_{\tau}}=10^{-1}$ eV, $m_{\nu_{\mu}}=8.8\cdot 10^{-2}$ eV, $m_{\nu_{e}}=8.8\cdot 10^{-2}$ eV
\end{itemize}

\subsection{Gauge couplings unification}
\begin{figure}[pt]
\begin{center} 
\epsfig{file=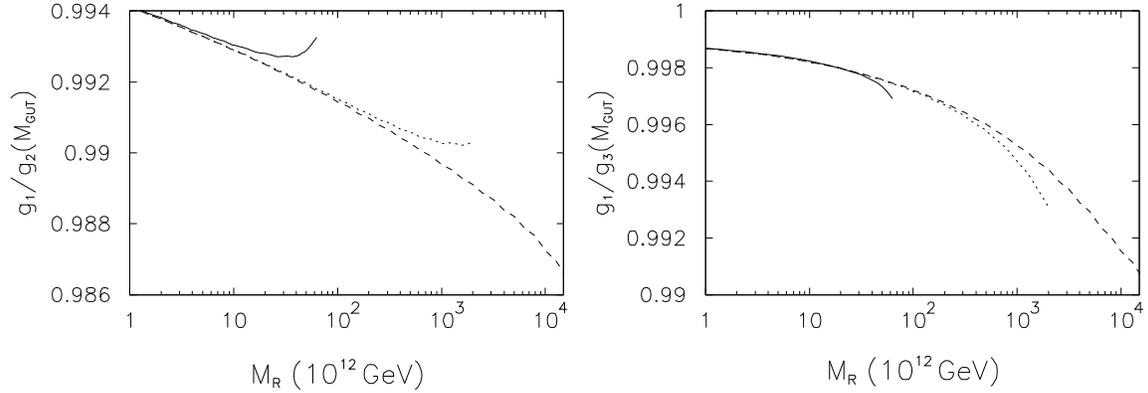,width=\linewidth}
\end{center}
\caption{The ratios $\frac{g_{1}}{g_{2}}(M_{GUT})$ and  $\frac{g_{1}}{g_{3}}(M_{GUT})$ as a function of $M_{R}$ for  $m_{\nu_{\tau}}=m_{\nu_{\mu}}=m_{\nu_{e}}=1$ eV (solid line), $m_{\nu_{\tau}}=10^{-1}$ eV, $m_{\nu_{\mu}}=8.8\cdot 10^{-2}$ eV, $m_{\nu_{e}}=8.8\cdot 10^{-2}$ eV (dashed line) and $m_{\nu_{\tau}}=4.5\cdot 10^{-2}$ eV, $m_{\nu_{\mu}}=10^{-3}$ eV, $m_{\nu_{e}}=10^{-4}$ eV (dotted line) .}\label{1}
\end{figure}  
In the Minimal Supersymmetric Standard Model exact unification of the gauge couplings at the energy $M_{GUT}=1.5\times 10^{16}$ GeV is obtained at the two-loop level for the supersymmetry breaking scale $M_{SUSY}=1$ TeV. Since the right-handed neutrinos are singlets under the MSSM gauge group, the unification of the gauge couplings is affected only by the presence of the Dirac neutrino Yukawa coupling at the two-loop level above the Majorana scale $M_{R}$. In Fig.\ref{1} we present the dependence of the ratios $g_{1}/g_{3}$ and $g_{1}/g_{2}$ in the MSSM at the GUT scale as a function of the Majorana mass for different sets of the neutrino masses at the electroweak scale. The effect is expected to be the stronger the larger is the initial value of $Y_{n}$, which is related through the see-saw mechanism to the light neutrino mass and the Majorana scale, $Y_{n}\sim \sqrt{m_{\nu}M_{R}}$. The lines are interrupted when neutrino Yukawa couplings become non-perturbative, i.e. $Y^{2}_{n}(M_{R})/4\pi\approx 1$. 

Unification of the gauge couplings is very weakly affected by the neutrino Yukawa couplings: the biggest effect is smaller than 1\%. It is not a surprise since $Y_{n}$ arise in $\beta(g)$ functions only as a two-loop correction. Our results agree with those presented in~\cite{casas}.

\subsection{$b-\tau$ unification}
\begin{figure}[pt] 
\begin{center}
\epsfig{file=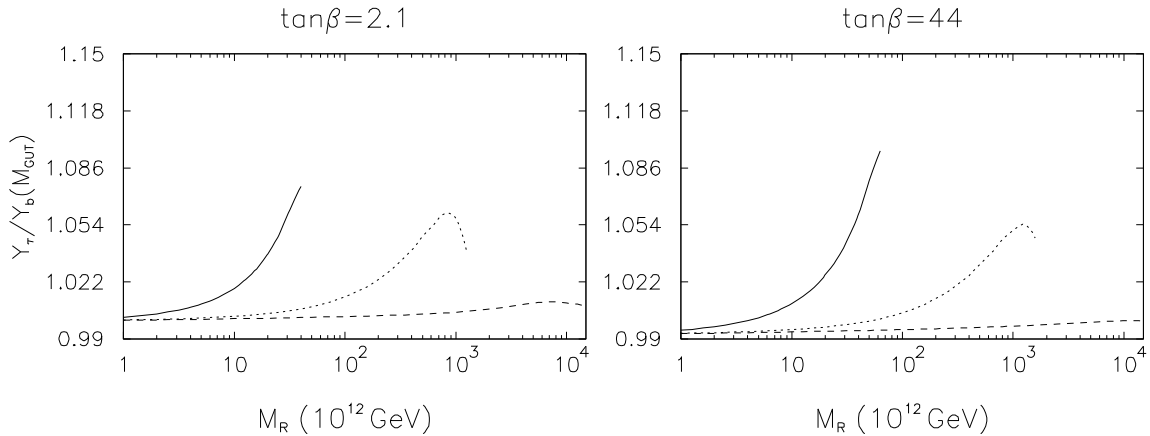, width=\linewidth}
\end{center}
\caption{The ratio $\frac{Y_{\tau}}{Y_{b}}(M_{GUT})$ as a function of $M_{R}$ for different values of $\tan\beta$  and $m_{\nu_{\tau}}=m_{\nu_{\mu}}=m_{\nu_{e}}=1$ eV (solid line), $m_{\nu_{\tau}}=10^{-1}$ eV, $m_{\nu_{\mu}}=8.8\cdot 10^{-2}$ eV, $m_{\nu_{e}}=8.8\cdot 10^{-2}$ eV (dashed line) and $m_{\nu_{\tau}}=4.5\cdot 10^{-2}$ eV, $m_{\nu_{\mu}}=10^{-3}$ eV, $m_{\nu_{e}}=10^{-4}$ eV (dotted line).}\label{2}
\end{figure}  
In the absence of massive neutrinos and with the initial values (\ref{input}), the $b-\tau$ unification is obtained for $\tan\beta=2.1$ and $\tan\beta=44$. For large values of $\tan\beta$ the finite supersymmetric one-loop correction to the bottom mass $m_{b}(M_{Z})$~\cite{radcor} becomes very important. The effects of threshold corrections introduces the dependence on the spectrum of the MSSM, however the loop correction to $m_{b}(M_{Z})$ is only weakly dependent on sparticle masses~\cite{susycor} and does not change after rescaling of masses of all SUSY particles. 

Contrary to the case of the gauge couplings, neutrinos affect the RGEs of Yukawa couplings at one-loop. Hence one can expect that their influence on the $b-\tau$ unification is much more significant. In Fig.\ref{2} we plot the ratio $Y_{\tau}/Y_{b}(M_{GUT})$ at the GUT scale as a function of $M_{R}$ for different values of the neutrino masses and $\tan\beta=2.1,44$ respectively. We fix supersymmetric parameters so that the bottom-tau unification holds for $\tan\beta=44$ and $M_{SUSY}=1$ TeV in the case of massless neutrinos. The lines are terminated when the neutrino Yukawa coupling reaches the non-perturbative region. 

\begin{figure}[t] 
\begin{center}
\epsfig{file=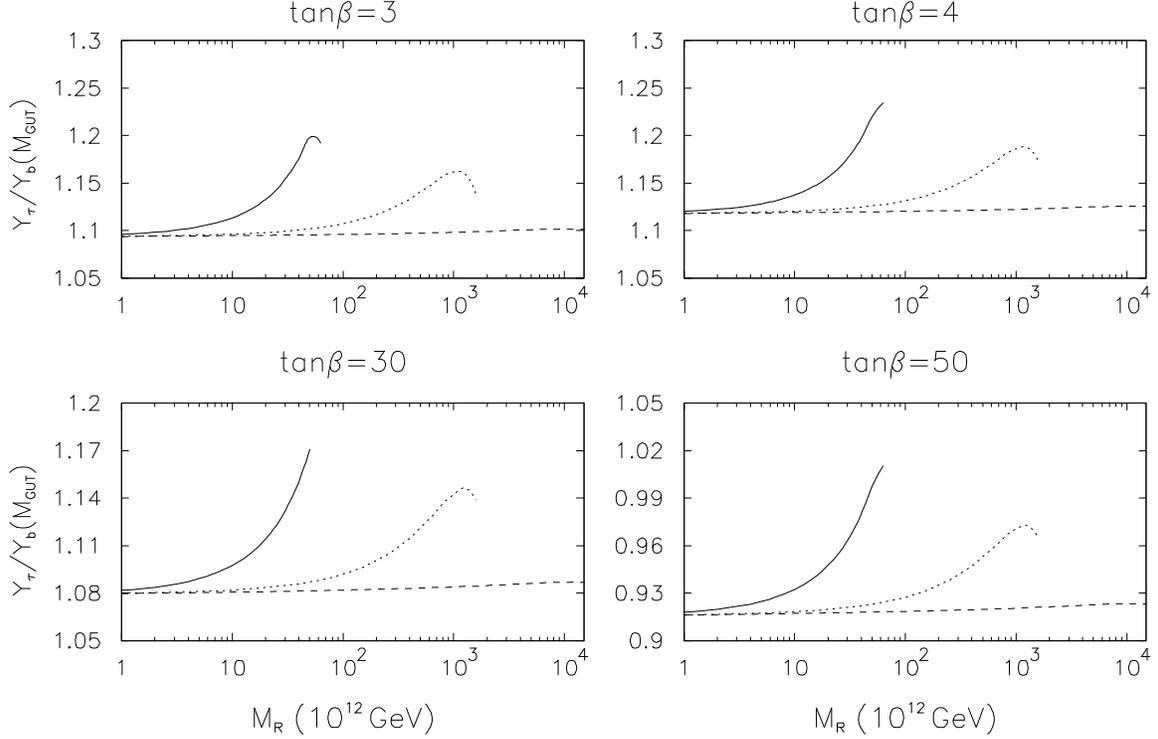,width=\linewidth}
\end{center}
\caption{The ratio $\frac{Y_{\tau}}{Y_{b}}(M_{GUT})$ as a function of $M_{R}$ for different values of $\tan\beta$  and $m_{\nu_{\tau}}=m_{\nu_{\mu}}=m_{\nu_{e}}=1$ eV (solid line), $m_{\nu_{\tau}}=10^{-1}$ eV, $m_{\nu_{\mu}}=8.8\cdot 10^{-2}$ eV, $m_{\nu_{e}}=8.8\cdot 10^{-2}$ eV (dashed line) and $m_{\nu_{\tau}}=4.5\cdot 10^{-2}$ eV, $m_{\nu_{\mu}}=10^{-3}$ eV, $m_{\nu_{e}}=10^{-4}$ eV (dotted line).}\label{3}
\end{figure}

One observes that introduction of massive neutrinos can change the ratio $Y_{\tau}/Y_{b}$ even by 12\%. As a result, the $b-\tau$ unification can hold for those values of $\tan\beta$ for which $Y_{\tau}/Y_{b}(M_{GUT})<1$ in the case of the massless neutrinos with the chosen  SUSY spectrum. To illustrate this effect, in Fig.\ref{3} we plot the ratio $Y_{\tau}/Y_{b}(M_{GUT})$ at the GUT scale as a function of $M_{R}$ for different values of the neutrino masses and $\tan\beta=3,4,30,50$ and with the supersymmetric parameters as in the previous case.
With massive neutrinos the ratio $Y_{\tau}/Y_{b}(M_{GUT})$ increases, hence  $b-\tau$ unification can be possible for $\tan\beta\lesssim 2.1$ and $\tan\beta\gtrsim 44$ for the chosen set of supersymmetric parameters. With other choice of SUSY parameters, unification can hold in the range of $\tan\beta$ given in~\cite{susycor}. However, addition of massive neutrinos cannot assure unification for $2.1\lesssim\tan\beta\lesssim 10$.  

For some parameters an extremum of the $Y_{\tau}/Y_{b}$ ratio is observed. Beyond this critical point the ratio $Y_{\tau}/Y_{b}$ decreases when the Majorana mass increases. This is due to the fact that the neutrino Yukawa coupling does not affect the bottom Yukawa $\beta$-function, but does enter the one of $\tau$ with positive sign. Therefore, the larger is the initial value of $Y_{n}$, the 'faster' is the running of the $\tau$ Yukawa coupling. However, for fixed $m_{\nu}$, larger $Y_{n}$ implies larger Majorana scale i.e. shorter the interval of scales at which the neutrino Yukawa coupling affects the running of $Y_{\tau}$. Therefore, for $M_{R}$ close to the GUT scale it can happen that, though $\tau$ Yukawa coupling is stronger renormalized than in smaller $M_{R}$ case, the final value of $Y_{\tau}$ will be smaller.

\begin{figure}[pt] 
\begin{center}
\epsfig{file=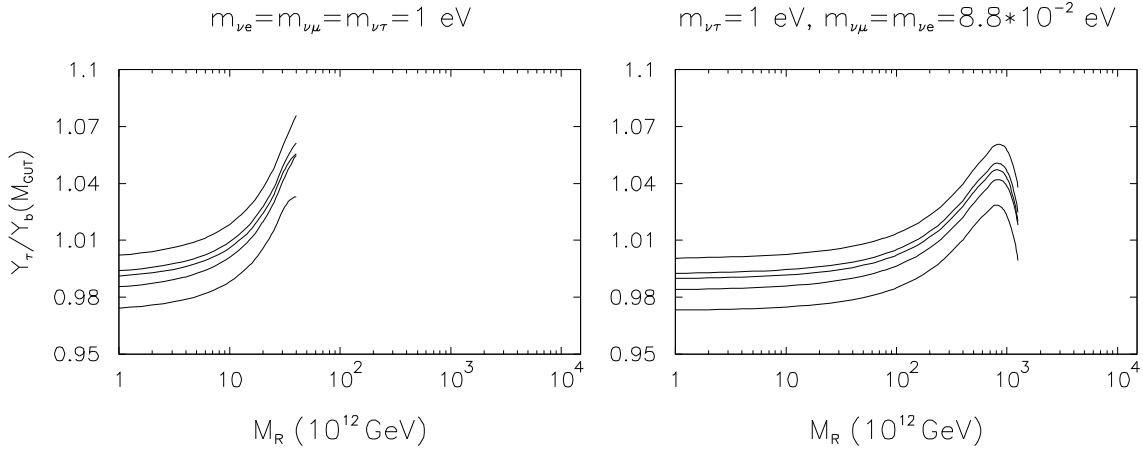,width=\linewidth}
\end{center}
\caption{The ratio $\frac{Y_{\tau}}{Y_{b}}(M_{GUT})$ as a function of $M_{R}$ for $\tan\beta=2.1$, different sets of neutrino masses and different values of supersymmetric threshold which, starting from below, read $M_{SUSY}=0.2,0.4,0.6,0.7,1$ TeV.}\label{5} 
\end{figure}

\begin{figure}[h] 
\begin{center}
\epsfig{file=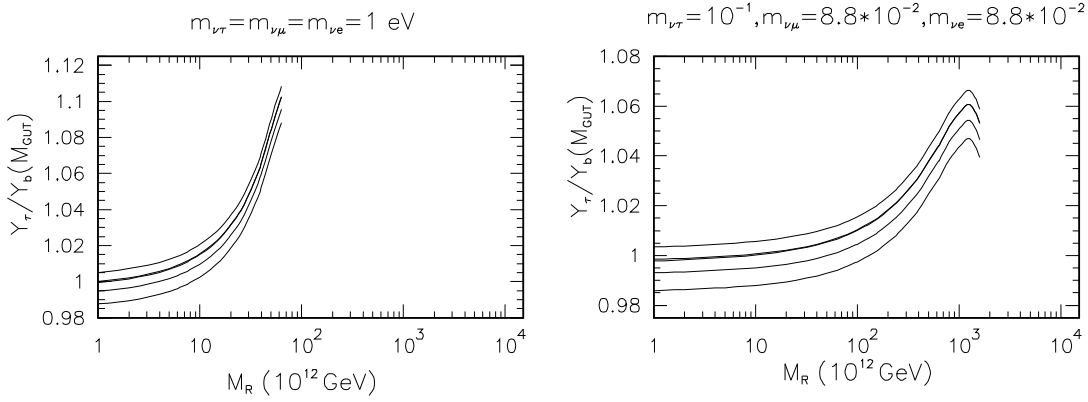, width=\linewidth}
\end{center}
\caption{The ratio $\frac{Y_{\tau}}{Y_{b}}(M_{GUT})$ as a function of $M_{R}$ for $\tan\beta=44$, different sets of neutrino masses and different values of supersymmetric threshold which, starting from below, read $M_{SUSY}=20,10,4,1$ TeV.}\label{6} 
\end{figure}

The other possibility to obtain $b-\tau$ unification in the presence of massive neutrinos is to keep $\tan\beta$ fixed while changing superparticle spectrum. The results of such an analysis are presented in Figs.\ref{5},\ref{6} for $\tan\beta=2.1$ and $\tan\beta=44$. As usually, we plot the ratio $Y_{\tau}/Y_{b}(M_{GUT})$ as a function of the Majorana scale for different values of supersymmetric threshold $M_{SUSY}$ and neutrino masses as in the previous case. 
In the large $\tan\beta$ case we have to pay attention to the one-loop correction to the bottom quark mass~\cite{radcor}. Since it depends on the particles masses, its magnitude might depend strongly on the $M_{SUSY}$ scale. However, it has been emphasized \cite{radcor} that this correction does not change after rescaling of masses of all SUSY particles, hence it only weakly depends on the superpartner masses.

One can see that for $\tan\beta=2.1$ the $b-\tau$ unification is possible only if the supersymmetry breaking scale is decreased. This is due to the fact that the Standard Model $\beta$-function of the $\tau$ Yukawa coupling decreases 'slower' than its supersymmetric counterpart. Thus, the smaller $M_{SUSY}$ scale, the smaller is the finite value of $Y_{\tau}$. At the same time the $b$ Yukawa coupling is more strongly renormalized by the MSSM RGEs than by the SM ones. However, this effect is very weak for small $\tan\beta$, hence $Y_{b}$ is almost unchanged by moving the superparticle decoupling scale. This way one can decrease the $Y_{\tau}/Y_{b}$ ratio by decreasing $M_{SUSY}$.

The situation is opposite for $\tan\beta=44$ case. In this case one has to choose larger values of $M_{SUSY}$ to get exact $b-\tau$ unification . It is possible even for supersymmetry breaking scale as high as 20 TeV. The mechanism is similar to the one in the previous case. However, this time the effects of changing $M_{SUSY}$ are stronger in the bottom quark Yukawa coupling case. The larger supersymmetric threshold, the larger is the final value of $Y_{b}$ which makes the ratio $Y_{\tau}/Y_{b}$ smaller.  

\section{Conclusions}
In this paper we have investigated the influence of neutrino masses on the unification of the gauge and Yukawa couplings in the Minimal Supersymmetric Standard Model. Neutrinos effect on the unification of the gauge couplings turned out to be very weak, as one could expect for the second-order correction. The ratios $g_{1}/g_{2}$ and $g_{1}/g_{3}$ at the GUT scale are changed by less than 0.7\%. 

The effect of the neutrino Yukawa couplings is much larger in the case of the $b-\tau$ unification. Generically the ratio $Y_{\tau}/Y_{b}$ increases in the presence of the right-handed neutrinos. The effect depends on the neutrino decoupling scale and on the light neutrino masses and can reach even 12\%. 

This mechanism can make the $b-\tau$ unification possible for those values of the MSSM parameters which were previously disfavored. It allows the superparticle decoupling scale to be smaller (bigger) than 1 TeV for $\tan\beta=2.1$ ($\tan\beta=44$). Moreover, it makes unification possible for those values of $\tan\beta$ for which $Y_{\tau}/Y_{b}<1$ at the GUT scale in the absence of massive neutrinos. However, this effect does not enlarge the range of $\tan\beta$, consistent with $b-\tau$ unification with the supersymmetric one-loop correction to the bottom quark mass included.

\section{Appendix}
Two-loop renormalization group equations of the Yukawa couplings above the Majorana scale:
\begin{eqnarray}
\beta(\mathbf{Y}_{u})=\frac{1}{16\pi^{2}}[Tr(3\mathbf{Y}_{u}\mathbf{Y}_{u}^{\dagger}+\mathbf{Y}_{n}\mathbf{Y}_{n}^{\dagger})+3\mathbf{Y}_{u}^{\dagger}\mathbf{Y}_{u}+\mathbf{Y}_{d}^{\dagger}\mathbf{Y}_{d}-(\frac{16}{3}g_{3}^{2}+3g_{2}^{2}+\frac{13}{15}g_{1}^{2})]\mathbf{Y}_{u}+
\nonumber\\
+(\frac{1}{16\pi^{2}})^{2}[-3Tr(3\mathbf{Y}_{u}\mathbf{Y}_{u}^{\dagger}\mathbf{Y}_{u}\mathbf{Y}_{u}^{\dagger}+\mathbf{Y}_{u}\mathbf{Y}_{u}^{\dagger}\mathbf{Y}_{d}\mathbf{Y}_{d}^{\dagger}+\mathbf{Y}_{n}\mathbf{Y}_{n}^{\dagger}\mathbf{Y}_{n}\mathbf{Y}_{n}^{\dagger})-\qquad\qquad\qquad\;\;\,
\nonumber\\
-\mathbf{Y}_{d}^{\dagger}\mathbf{Y}_{d}Tr(3\mathbf{Y}_{d}\mathbf{Y}_{d}^{\dagger}+\mathbf{Y}_{e}\mathbf{Y}_{e}^{\dagger})-3\mathbf{Y}_{u}\mathbf{Y}_{u}^{\dagger}Tr(3\mathbf{Y}_{u}\mathbf{Y}_{u}^{\dagger}+\mathbf{Y}_{n}\mathbf{Y}_{n}^{\dagger})-\qquad\quad\,
\nonumber\\
-4\mathbf{Y}_{u}\mathbf{Y}_{u}^{\dagger}\mathbf{Y}_{u}\mathbf{Y}_{u}^{\dagger}-2\mathbf{Y}_{d}\mathbf{Y}_{d}^{\dagger}\mathbf{Y}_{d}\mathbf{Y}_{d}^{\dagger}-2\mathbf{Y}_{d}^{\dagger}\mathbf{Y}_{d}\mathbf{Y}_{u}^{\dagger}\mathbf{Y}_{u}+\frac{2}{5}g_{1}^{2}\mathbf{Y}_{d}\mathbf{Y}_{d}^{\dagger}+\qquad\;\;\:\,
\nonumber\\
+(\frac{2}{5}g_{1}^{2}+6g_{2}^{2})\mathbf{Y}_{u}\mathbf{Y}_{u}^{\dagger}+(16g_{3}^{2}+\frac{4}{5}g_{1}^{2})Tr\mathbf{Y}_{u}\mathbf{Y}_{u}^{\dagger}-\frac{16}{9}g_{3}^{4}+8g_{3}^{2}g_{2}^{2}+\qquad\;\:\,
\nonumber\\
+\frac{136}{45}g_{3}^{2}g_{1}^{2}+\frac{15}{2}g_{2}^{4}+g_{2}^{2}g_{1}^{2}+\frac{2743}{450}g_{1}^{4}]\mathbf{Y}_{u}\qquad\qquad\qquad\qquad\qquad\qquad\quad\:
\end{eqnarray}
\begin{eqnarray}
\beta(\mathbf{Y}_{d})=\frac{1}{16\pi^{2}}[Tr(3\mathbf{Y}_{d}\mathbf{Y}_{d}^{\dagger}+\mathbf{Y}_{e}\mathbf{Y}_{e}^{\dagger})+3\mathbf{Y}_{d}^{\dagger}\mathbf{Y}_{d}+\mathbf{Y}_{u}^{\dagger}\mathbf{Y}_{u}-(\frac{16}{3}g_{3}^{2}+3g_{2}^{2}+\frac{7}{15}g_{1}^{2})]\mathbf{Y}_{d}+
\nonumber\\
+(\frac{1}{16\pi^{2}})^{2}[-3Tr(3\mathbf{Y}_{d}\mathbf{Y}_{d}^{\dagger}\mathbf{Y}_{d}\mathbf{Y}_{d}^{\dagger}+\mathbf{Y}_{u}\mathbf{Y}_{u}^{\dagger}\mathbf{Y}_{d}\mathbf{Y}_{d}^{\dagger}+\mathbf{Y}_{e}\mathbf{Y}_{e}^{\dagger}\mathbf{Y}_{e}\mathbf{Y}_{e}^{\dagger})-\qquad\qquad\qquad\;\;\;\:
\nonumber\\
-3\mathbf{Y}_{d}^{\dagger}\mathbf{Y}_{d}Tr(3\mathbf{Y}_{d}\mathbf{Y}_{d}^{\dagger}+\mathbf{Y}_{e}\mathbf{Y}_{e}^{\dagger})-\mathbf{Y}_{u}\mathbf{Y}_{u}^{\dagger}Tr(3\mathbf{Y}_{u}\mathbf{Y}_{u}^{\dagger}+\mathbf{Y}_{n}\mathbf{Y}_{n}^{\dagger})-\qquad\;\,
\nonumber\\
-4\mathbf{Y}_{d}\mathbf{Y}_{d}^{\dagger}\mathbf{Y}_{d}\mathbf{Y}_{d}^{\dagger}-2\mathbf{Y}_{u}\mathbf{Y}_{u}^{\dagger}\mathbf{Y}_{d}\mathbf{Y}_{d}^{\dagger}-2\mathbf{Y}_{u}^{\dagger}\mathbf{Y}_{u}\mathbf{Y}_{u}^{\dagger}\mathbf{Y}_{u}+\frac{4}{5}g_{1}^{2}\mathbf{Y}_{u}\mathbf{Y}_{u}^{\dagger}+\qquad\;\:
\nonumber\\
+(\frac{4}{5}g_{1}^{2}+6g_{2}^{2})\mathbf{Y}_{d}\mathbf{Y}_{d}^{\dagger}+(16g_{3}^{2}-\frac{2}{5}g_{1}^{2})Tr\mathbf{Y}_{d}\mathbf{Y}_{d}^{\dagger}+\frac{6}{5}g_{1}^{2}Tr(\mathbf{Y}_{e}\mathbf{Y}_{e}^{\dagger})-\qquad\,
\nonumber\\
-\frac{16}{9}g_{3}^{4}+8g_{3}^{2}g_{2}^{2}+\frac{8}{9}g_{3}^{2}g_{1}^{2}+\frac{15}{2}g_{2}^{4}+g_{2}^{2}g_{1}^{2}+\frac{287}{90}g_{1}^{4}]\mathbf{Y}_{d}\qquad\qquad\qquad\quad\;\:
\end{eqnarray}
\begin{eqnarray}
\beta(\mathbf{Y}_{n})=\frac{1}{16\pi^{2}}[Tr(3\mathbf{Y}_{u}\mathbf{Y}_{u}^{\dagger}+\mathbf{Y}_{n}\mathbf{Y}_{n}^{\dagger})+3\mathbf{Y}_{n}^{\dagger}\mathbf{Y}_{n}+\mathbf{Y}_{e}^{\dagger}\mathbf{Y}_{e}-(3g_{2}^{2}+\frac{3}{5}g_{1}^{2})]\mathbf{Y}_{n}+\qquad\quad\;
\nonumber\\
+(\frac{1}{16\pi^{2}})^{2}[-3Tr(3\mathbf{Y}_{u}\mathbf{Y}_{u}^{\dagger}\mathbf{Y}_{u}\mathbf{Y}_{u}^{\dagger}+\mathbf{Y}_{n}\mathbf{Y}_{n}^{\dagger}\mathbf{Y}_{n}\mathbf{Y}_{n}^{\dagger}+\mathbf{Y}_{n}\mathbf{Y}_{n}^{\dagger}\mathbf{Y}_{e}\mathbf{Y}_{e}^{\dagger})-\qquad\qquad\qquad\,
\nonumber\\
-3\mathbf{Y}_{n}^{\dagger}\mathbf{Y}_{n}Tr(3\mathbf{Y}_{u}\mathbf{Y}_{u}^{\dagger}+\mathbf{Y}_{n}\mathbf{Y}_{n}^{\dagger})-3\mathbf{Y}_{e}\mathbf{Y}_{e}^{\dagger}Tr(3\mathbf{Y}_{d}\mathbf{Y}_{d}^{\dagger}+\mathbf{Y}_{e}\mathbf{Y}_{e}^{\dagger})-\qquad\:
\nonumber\\
-4\mathbf{Y}_{n}\mathbf{Y}_{n}^{\dagger}\mathbf{Y}_{n}\mathbf{Y}_{n}^{\dagger}-2\mathbf{Y}_{e}\mathbf{Y}_{e}^{\dagger}\mathbf{Y}_{n}\mathbf{Y}_{n}^{\dagger}-2\mathbf{Y}_{e}^{\dagger}\mathbf{Y}_{e}\mathbf{Y}_{e}^{\dagger}\mathbf{Y}_{e}+\qquad\qquad\qquad\quad\;\:\:\,
\nonumber\\
+\frac{6}{5}g_{1}^{2}Tr(Y_{n}Y_{n}^{\dagger})+6g_{2}^{2}\mathbf{Y}_{n}\mathbf{Y}_{n}^{\dagger}+(16g_{3}^{2}+\frac{4}{5}g_{1}^{2})Tr\mathbf{Y}_{u}\mathbf{Y}_{u}^{\dagger}+\qquad\qquad\qquad,
\nonumber\\
+\frac{15}{2}g_{2}^{4}+\frac{9}{5}g_{2}^{2}g_{1}^{2}+\frac{207}{50}g_{1}^{4}]\mathbf{Y}_{n}\qquad\qquad\qquad\qquad\qquad\qquad\qquad\qquad\quad\;\label{yuk}
\end{eqnarray}
\begin{eqnarray}
\beta(\mathbf{Y}_{e})=\frac{1}{16\pi^{2}}[Tr(3\mathbf{Y}_{d}\mathbf{Y}_{d}^{\dagger}+\mathbf{Y}_{e}\mathbf{Y}_{e}^{\dagger})+3\mathbf{Y}_{e}^{\dagger}\mathbf{Y}_{e}+\mathbf{Y}_{n}^{\dagger}\mathbf{Y}_{n}-(3g_{2}^{2}+\frac{9}{5}g_{1}^{2})]\mathbf{Y}_{e}+\qquad\quad\;
\nonumber\\
+(\frac{1}{16\pi^{2}})^{2}[-3Tr(3\mathbf{Y}_{d}\mathbf{Y}_{d}^{\dagger}\mathbf{Y}_{d}\mathbf{Y}_{d}^{\dagger}+\mathbf{Y}_{u}\mathbf{Y}_{u}^{\dagger}\mathbf{Y}_{d}\mathbf{Y}_{d}^{\dagger}+\mathbf{Y}_{e}\mathbf{Y}_{e}^{\dagger}\mathbf{Y}_{e}\mathbf{Y}_{e}^{\dagger}+\mathbf{Y}_{e}\mathbf{Y}_{e}^{\dagger}\mathbf{Y}_{n}\mathbf{Y}_{n}^{\dagger})-\qquad\,
\nonumber\\
-3\mathbf{Y}_{e}^{\dagger}\mathbf{Y}_{e}Tr(3\mathbf{Y}_{d}\mathbf{Y}_{d}^{\dagger}+\mathbf{Y}_{e}\mathbf{Y}_{e}^{\dagger})-3\mathbf{Y}_{n}\mathbf{Y}_{n}^{\dagger}Tr(3\mathbf{Y}_{u}\mathbf{Y}_{u}^{\dagger}+\mathbf{Y}_{n}\mathbf{Y}_{n}^{\dagger})-\qquad\:
\nonumber\\
-4\mathbf{Y}_{e}\mathbf{Y}_{e}^{\dagger}\mathbf{Y}_{e}\mathbf{Y}_{e}^{\dagger}-2\mathbf{Y}_{n}\mathbf{Y}_{n}^{\dagger}\mathbf{Y}_{e}\mathbf{Y}_{e}^{\dagger}-2\mathbf{Y}_{n}^{\dagger}\mathbf{Y}_{n}\mathbf{Y}_{n}^{\dagger}\mathbf{Y}_{n}+\qquad\qquad\qquad\quad\;\:\:\,
\nonumber\\
+\frac{6}{5}g_{1}^{2}Tr(Y_{e}Y_{e}^{\dagger})+6g_{2}^{2}\mathbf{Y}_{e}\mathbf{Y}_{e}^{\dagger}+(16g_{3}^{2}-\frac{2}{5}g_{1}^{2})Tr\mathbf{Y}_{d}\mathbf{Y}_{d}^{\dagger}+\qquad\qquad\qquad\,
\nonumber\\
+\frac{15}{2}g_{2}^{4}+\frac{9}{5}g_{2}^{2}g_{1}^{2}+\frac{27}{2}g_{1}^{4}]\mathbf{Y}_{e}\qquad\qquad\qquad\qquad\qquad\qquad\qquad\qquad\quad\quad
\end{eqnarray}


\begin{thebibliography}{99}
\bibitem{unif}
H. Georgi, H. Quinn, S. Weinberg, {\sl Phys. Rev. Lett.} \textbf{33}, 451, (1974);

H. Georgi, S. L. Glashow, {\sl Phys. Rev. Lett.} \textbf{32}, 438, (1974).

\bibitem{susycor}
P. Chankowski, J. Ellis, M. Olechowski, S. Pokorski, {\sl Nucl. Phys} \textbf{544}, 39, (1999), preprint CERN-TH-98-119 (hep-ph/9808275).

\bibitem{unsm}
U. Amaldi, W. de Boer, H. F\"urstenau, {\sl Phys. Lett.} \textbf{B260}, 447, (1991).

\bibitem{sk}
Y. Fukuda \textit{et al}., Super-Kamiokande Collaboration, {\sl Phys. Rev. Lett.} \textbf{81}, 1562, (1998), {\sl Phys. Lett.} \textbf{B433}, 9, (1998);

S. Hatakeyama \textit{et al}., Kamiokande Collaboration, {\sl Phys. Rev. Lett.} \textbf{81}, 2016, (1998).

\bibitem{chooz}
M. Apollonio \textit{et al}., CHOOZ Collaboration, {\sl Phys. Lett.} \textbf{420}, 397, (1998).
\bibitem{solar}
Y. Fukuda \textit{et al}., Super-Kamiokande Collaboration, {\sl Phys. Rev. Lett.} \textbf{82}, 1810, (1999);

W. Hampel \textit{et al}., GALLEX Collabortation, {\sl Phys. Lett.} \textbf{B477}, 127, (199).
\bibitem{vacos}
V. N. Gribov and B. M. Pontecorvo, {\sl Phys. Lett.} \textbf{B28}, 493, (1969).
\bibitem{msw}
L. Wolfenstein, {\sl Phys. Rev.} \textbf{D17}, 2369, (1978);

S. Mikheyev and A. Smirnov, {\sl Sov. J. Nucl. Phys.} \textbf{42}, 913, (1985).

\bibitem{see-saw}
R. Mohapatra and G. Sejanovic, {\sl Phys. Rev. Lett.} \textbf{44}, 912, (1980), {\sl Phys. Rev.} \textbf{D23}, 165, (1981).

\bibitem{casas}
J. A. Casas, J. R. Espinosa, A. Ibarra, I. Navarro, {\sl Phys. Rev.} \textbf{D63}, (2001).

\bibitem{Carena}
M. Carena, J. Ellis, S. Lola and C.E.M. Wagner, {\sl Eur. Phys. J.} \textbf{C12}, 507 (2000), preprint CERN-TH/99-173.

\bibitem{smrge}
M. E. Machacek, M. T. Vaughn, {\sl Nucl.Phys.} \textbf{B236}, 221, (1984).

\bibitem{rge}
S. P. Martin and M. T. Vaughn, preprint NUB-TH-3081 (hep-ph/9311340).

\bibitem{ratz}
S. Antusch and M. Ratz, preprint TUM-HEP-453/01 (hep-ph/0203027). 

\bibitem{radcor}
M. Carena, M. Olechowski, S. Pokorski, C.E.M. Wagner, {\sl Nucl. Phys.} \textbf{B426}, 269, (1994).

\end{thebibliography}
\end{document}